# Using Collaborative Visual Analytics for Innovative Industry-inspired Learning Activities

Olivera Marjanovic
Business Information Systems
University of Sydney Business School
Sydney, NSW 2006 Australia
Email: olivera.marjanovic@sydney.edu.au

## Abstract

Inspired by leading industry practices, this paper describes an innovative learning activity designed to combine data visualisation and cross-functional collaboration supported by enterprise social media. The activity is structured around sharing, co-creation and negotiation of departmental/disciplinary insights across data silos, using both internal and external data. In addition to giving students access to state-of-the-art tools for visualisation (SAS-VA) and collaboration (Yammer), an even more important educational objective is to expose students to the complexities of deriving information (i.e. meaning) from enterprise-wide (meaning-free) data stored by business intelligence & analytics systems. This human-driven and human-centric process of making sense of data in context requires collaboration across functional silos, especially when dealing with complex multi-disciplinary challenges. Starting from an industry-informed business scenario, the paper describes the main steps of an innovative data visualisation and collaboration activity, discusses possible alternative software platforms and offers some ideas for the future work.

## 1 Introduction

Worldwide proliferation of the so-called "Big data" – large *volumes* of real-time data (*velocity*) coming from a *variety* of sources such as social media, mobile devices, and sensors - is creating new challenges and opportunities across all industry sectors. Consequently, reporting and simple analytical tools are no longer sufficient to support today's complex decision-making needs. Especially, when it comes to data discovery (La Valle et al. 2010) or "story telling with data" (Davenport 2103; Kosara and Mackinlay 2013; Naidoo 2013). "Knowing what happened and why it happened are no longer adequate. Organizations need to know what is happening now, what is likely to happen next and what actions should be taken to get optimal results" (La Valle et al. 2010, p.22). Data visualisation and "story telling with data" have been identified as the most important future trends in Business Intelligence and Analytics (BI&A) (Stoder 2013; HBR Insight Center 2013). Consequently, the new term "Visual Analytics"(VA) has been coined by industry to describe data visualisation combined with new practices for test-and-learn-inquiry and data discovery (Stodder 2013) as well as to distinguish VA from the mainstream BI&A that remains focused on numerical data.

In spite of a rapidly growing amount of data and more sophisticated BI&A tools, many organizations are still looking for better ways to obtain value from their data (LaValle et al. 2010, p. 22). Both industry and research communities now recognise the challenge of deriving *meaningful and timely insights (information)* from *data* to be greater than ever. "However, despite the hopes of many, insights do not emerge automatically out of mechanically applying analytical tools to data" (Sharma, Mithas and Kankankalli 2014, p.25).

A process of deriving information from *meaning-free* data contained in BI&A systems, in order to take a value-adding action, is very complex for most organisations and involves multiple actors from different parts of the organisation (Sharma et al. 2014). Typically, data resides in functional silos. When propagated across different functional and hierarchical boundaries, data are taken from the original context and then subjected to different interpretations, contexts and perspectives - often in direct conflict. Yet, complex problem solving requires collaboration where different data-informed "silo" perspectives are explored through collective "sense-making" in order to agree on a preferred action. While different collaborative tools have been used for decades, the challenge of *making sense* *of enterprise-wide data through collaboration* across functional silos provides a new (BI&A) context for their organisational application.

Current industry trends have resulted in an unprecedented demand for BI&A professionals, as confirmed by prominent industry reports (La Valle et al. 2010; Manyika et al. 2011), as well as by three international surveys of educators of Business Intelligence, Business Analytics or related fields (Wixom et al. 2010; Wixom and Ariyachandra 2011; Wixom et al. 2014). In response, universities world-wide are rapidly increasing the number and variety of BI&A-related courses, programs and degrees they offer. "Business schools have been challenged to increase business analytics skills across all of their degree areas including supply chain management, accounting, marketing and finance" (Rodammer, Speier-Pero and Haan 2015, p.1). For example, a recent survey by Wixom et al. (2014) identified 131 full-time university degrees directly related to BA in 2012 compared to only 15 in the previous 2010 survey. Among new programs, 47 are being offered at the undergraduate level, compared to only 3 just 2 years earlier. While many university offerings remain focused on more traditional BI&A and numerical data, the latest report by Gupta, Goul and Dinter (2015) on a multi-year development of the model curricula for BI, BA and BI&A in Business School Undergraduate, MS Graduate and MBA programs, recommends data visualisation to be included at all levels.

However, university educators are faced with the ongoing challenge of creating realistic learning activities, informed by industry practices (Wixom et al. 2014). Even more, focusing on technology (i.e. skill-based training) is not what industry wants. "Certifying students in real-world software is not the answer" (Wixom et al. 2014, p.10). In fact, when asked about their expectations and recommendations for BI&A educators, out of over 400 employers surveyed by Wixom et al. (2014) only 15 suggested that technical or vendor certification matter but only as an added bonus rather than a job requirement. What they recommend, and educators also need, are relevant and realistic learning experiences, that are currently lacking (Wixom et al. 2014; Gupta et al. 2015).

Informed by the above described industry trends, in this paper we describe an innovative teaching and learning activity designed to extend data visualisation with collaboration structured around sharing, co-creation and negotiation of departmental/disciplinary insights, using both internal and external sources of data. The main educational objective of this activity is to expose students to current industry practices and challenges of deriving insights (information) from data residing in organisational silos and then collaborating across those silos to make sense of the combined enterprise-wide data, in order to agree on a value-adding action. In terms of educational resources, this activity combines the existing SAS-VA platform currently (freely) available on the Teradata University Network portal (www.teradatauniversitynetwork.com) with the leading enterprise social media platform called Yammer. With both platforms being well-known industry leaders in their own (individual) application domains, in practical terms this activity provides students with an opportunity to use state-of-the-art tools for visualisation of very large data sets (in this case with 35 million records) and enterprise social media collaboration. The resulting activity was designed within a much larger project aiming to "translate" practitioner stories and experiences in using visual analytics into innovative industry-informed learning activities for students, described in (Marjanovic 2014a).

The paper is organized as follows. Section 2 offers a brief overview of data visualisation as it is currently discussed by industry and academia and motivates the need for cross-functional collaboration when making sense of enterprise-wide data. Section 3 describes the overall project that provides the main motivation and background for the design of the learning activity presented in this paper, as well as the research method. This is followed by a high-level overview of the main steps (stages) of the data visualization and collaboration activity (Section 4), described in more detail in Section 5. Section 6 positions the described activity in the context of Big data and discusses the alternative software platforms. Section 7 describes the current limitations of this work and provides some insights into our current and future work in this area.

## 2   Related work

Various industry thought leaders argue that one of the best ways to make sense of data (i.e. "quickly consume data") is through data visualisation (HBR Insight Center 2013; Sallam, Tapadinhas, Parenteau et al. 2014). "Data visualization is taking hold now because of two trends. The first: Big data is here, it must be analysed, and one of the best ways to make sense of it is with visual representations. The second: The tools to create good data visualizations are being democratized" (HBR Insight Center 2013, p1).

Even more, improving data visualisation and visual analysis for nontechnical users is one of the key recommendations by the 2013 TDWI Best Practice Report on Data Visualization and Discovery, made on the basis of their international survey of 453 companies, see (Stodder 2013). Nontechnical (e.g. business) users are now seen as critical for an organisation's ability to generate data-driven insights i.e. to derive useful information from data (facts). This is because business decision-makers have the necessary subject-matter expertise and experience to *make sense of data in context*, rather than delegate this task to data-scientists or IT experts who are not domain experts.

The latest development in visual analytics is best described as "story telling with visual data". More precisely, vendors and industry leaders call for new data visualisation practices of using visual analytics to tell stories and for new tools that are more suitable for business decision makers who don't normally have skills in "deep analytics" (Davenport 2013; Naido 2013; Kosara and Mackinlay 2013). "We have grown up on pie charts and bar charts, but there are probably at least tens, if not hundreds of alternative approaches to visual analytics. Narratives are a pretty good way to convey information in the past, so maybe we should be converting our data and analysis into stories… we've got a long way to go in terms of doing a better job of that" (Davenport 2013, p.1).

Yet, in spite of widespread awareness, current industry practices are still limited. "There are still lots of companies that don't have visualisation solutions in place, or if they do, it's only for a small number of people… but demand is growing" (Eckerson and Hammond 2011, p.5). The greatest concern is employees' very limited knowledge and skills in this area to enable them to make effective use of the already available tools (Stodder 2013, p.5). The skill-shortage in VA is expected to worsen (Eckerson 2011), creating even more challenges for organisations wanting to bring VA solutions to a wider range of business users, beyond data scientists and IT professionals. In turn, these industry trends are likely to create an even bigger demand for VA-related courses, especially among non-BI&A specialists.

Furthermore, regardless of VA tools being used, complex problems need to be examined from different perspectives, typically by decision-makers from different functional areas and/or different professional backgrounds. Often, they need to collaborate in order to combine their individual (limited) insights into a more holistic perspective and then decide on the most appropriate action. In fact, the importance of collaboration for deriving insights from data has been confirmed by both research and practice. "With more data in the hands of more people – and easier access to easy-to-use analytics – conversations about data and results from data analysis are happening more often. And becoming more important. And expected. So it's not surprising that improved collaboration is one of the most common organizational goals" (Bailey, Fu, Luppi et al. 2014).

Different collaborative tools have been used for many decades, ranging from group decision support systems (GDSS) and groupware of the 80s, to more recent enterprise social media platforms. While the topic of enterprise collaboration has been investigated quite extensively and for many years by many research communities, this project places collaboration in the context of BI&A. However, rather than focusing on collaboration, the primary focus is on BI&A, in particular *organisational challenges* of deriving insights from enterprise-wide data, through collective sense-making across functional silos.

The next section describes our research method of creating relevant and realistic learning activities, followed by an example of such an activity that was designed for business rather than IT or data-science students.

## 3　Research Method

The activity presented in this paper was developed within a much larger project focused on innovative teaching in VA (Marjanovic 2014a). One of the key objectives of the larger project was to 'translate' the leading industry practices (practitioner stories) in VA into innovative learning activities that resemble real-life challenges and practices of decision makers at all organizational levels (Marjanovic, 2014a). This section presents a brief overview of the research method used to derive educational activities from practitioner stories (as shown by Figure 1) that was also used when designing the activity presented later in this paper, for more details see (Marjanovic 2014a).

Inspired by the pressing need to offer a relevant and realistic learning experience to business students in BI&A, as indicated by Wixom et al. (2014), we started this project by focusing on the leading industry practices in VA. In the first step, we downloaded a large number of exemplary customer-success stories (65 industry cases) from vendors' web sites (SAS VA and Tableau). We selected these particular vendor companies as widely recognized analytical leaders in visual analytics. For example, the latest 2014 Gartner's Magic Quadrant report on BI&A platforms, positions both companies in the Leaders and Visionaries quadrant (Sallam et al. 2014).

After the customer success stories were collected, we proceeded to analyse them using the following research questions as lens:

- Who is using VA and for what purpose?

- What types of decisions do they make?

- What are their current challenges related to VA?

- What are the new opportunities created by VA?

- What are the required skills for using VA?

The main idea behind focusing on decision makers using VA is to put students into similar, industry-informed roles and in this way attempt to create a more realistic learning experience.

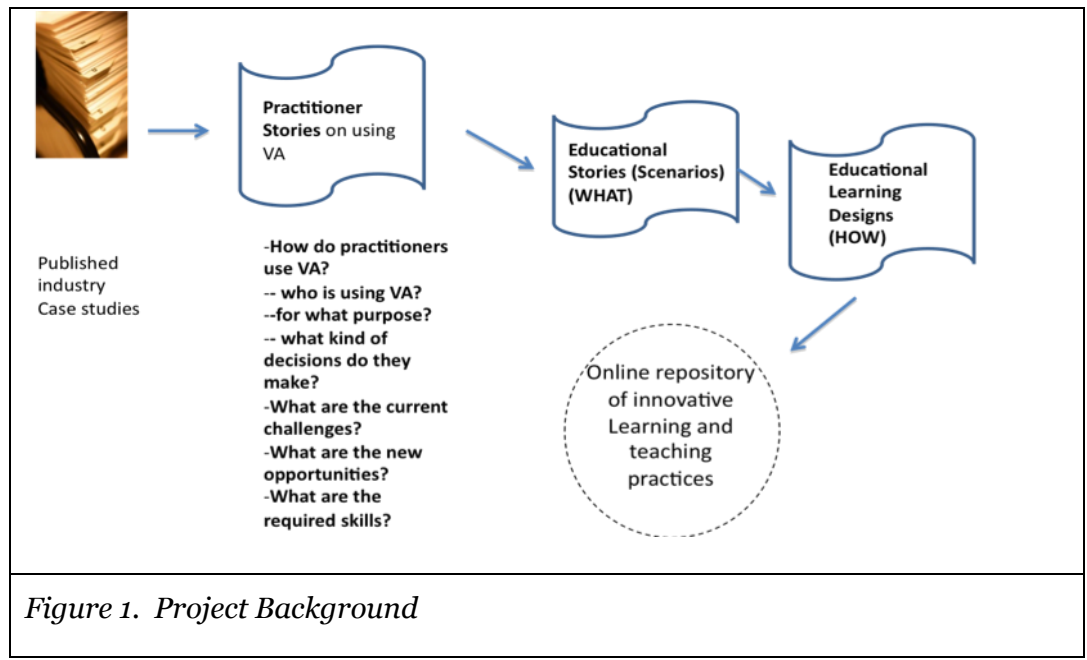

*Figure 1. Project Background*

The adopted method of data collection by downloading and analysing customer success stories from the vendors' web sites, was previously proposed and used by prominent BI&A researchers, (see Seddon and Constantinidis 2012; Seddon, Calvert and Yang, 2010) to study for example, the ways business analytics contribute to business value.

After these cases were analysed, the next challenge was to turn the obtained insights into industry-informed classroom activities. We argue that it is not realistic to aim to re-enact industry stories in a classroom environment, due to organizational complexities involved, as well as the normative aspects of these roles (e.g. their professional responsibility for possible outcomes of their actions). Instead, we decided to focus on underlying patterns of challenges and practices the leading VA practitioners are currently experiencing, based on the featured case studies. The main idea behind this is to expose students to the same types of VA-related challenges through simulated activities.

Through this process of looking for underlying patterns, we inferred some common types of challenges that could be found across different case studies and industry sectors, such as:
1. Organizational (functional) data silos leading to limited (functional) perspectives
2. Different sources of data, often in spreadsheets and functional databases
3. Data quality issues such as semantic and syntactic inconsistencies caused by for example, different sources of data and the associated meta-data
4. Collaboration required to combine different perspectives, especially when analysing complex problems and proposing a cause of action

As illustrated by Figure 1, these types of challenges were then incorporated into educational stories – hypothetical but realistic, industry-informed business scenarios. For example, we created scenarios about business challenges that in reality require different functional perspectives, often caused by organisational silos, leading to different interpretations of data, data inconsistencies and so on. The activity described in this paper illustrates one of these business scenarios.

While the previous step completes the process of "translating" practitioner stories into educational scenarios that could be used in a classroom, in this project we went further. Inspired by an objective to enable knowledge sharing among educators, we proceeded to capture these educational stories in a systematic form using a high-level model of learning designs (LDs) founded in the LD theory by Koper and Tattersall (2005). The resulting LDs are meant to give ideas and guidance (from educational perspective) rather than prescribe what to do.

Finally, as depicted by Figure 1, LDs are stored in an open online repository of teaching practices, implemented in wikispaces. Details of conceptual modelling of LDs using the LD theory and the LDs repository are out of scope of this paper and are presented elsewhere, see Marjanovic (2014b).

In summary, the above-described process (depicted by Figure 1) resulted in design of a number of innovative learning activities. The reminder of this paper will focus on one of these activities.

## 4    Combining Visual Analytics and Enterprise Social Media

The main idea for this activity came from a new industry trend of combining visual analytics and collaboration, identified and analysed through the above described research method. In fact, this particular trend is now recognized as one of the most important future areas of visual analytics, as argued by Bailey et al. (2014).

Thus, using industry-based cases as depicted by Figure 1, we identified a pattern of situations that require sharing of analytical insights across functional silos and collaboration to bring different perspectives together in order to better understand a particular business problem. Although collaboration was recognised as very important, the industry cases we analysed did not provide sufficient details about the organisational challenges of sharing VA-enabled insights across functional, hierarchical, organizational and even professional boundaries.

This particular limitation has led to an idea to combine VA with collaborative activities focused on sharing VA-enabled insights, deemed necessary in order to convert these insights into value-adding actions. From the educational perspective this learning activity was designed to show how sharing visual insights via an enterprise social media platform might reduce the impact of functional silos on decision-making or in some cases, create an even greater challenge.

The resulting activity was implemented by combining two leading industry platforms for data visualization (SAS VA) and enterprise social media-enabled collaboration (Yammer). We used the currently available cloud-based version of SAS VA, hosted by Teradata University Network (TUN) (www.teradatauniversitynetwork.com). For the collaborative part, we used a corporate (industrial-strength) version of the enterprise social-media platform Yammer (www.yammer.com), often described as the leading enterprise (corporate) Facebook-like web 2.0 platform (Chacos 2012). The main reason for selection of these platforms was their availability and cloud-based access making them maintenance free, thus easy to deploy in a classroom.

The resulting innovative activity was designed to include the following main steps (phases):

Step 1: Data visualisation using SAS VA (from Teradata University Network)

Step 2: Importing the resulting SAS VA visualisations into Yammer

Step 3: Collaborative activity based on a single (functional) perspective

Step 4: Collaborative activity of exploring multiple (disciplinary) perspectives, using internal and external social media data.

Step 5: Collaborative activity of translating VA-enabled insights into action

Step 6: Reflection-on-action (individual and/or collaborative)

The activity was designed to use one of the existing data sets provided by SAS-VA (on TUN). The activity also incorporated social media data (from Twitter and Google Trends) and in this way give students exposure to a wide variety of unstructured and external data sources and types. The following section describes these steps in more details.

# 5  Innovative Learning activity – The Main Steps

*- Business Scenario*

Insight Toy Company is an international company that manufactures and sells toys to retails (i.e. resellers "vendors") all over the world (SAS VA, 2014). In this context, students are asked to assume the roles of different functional managers and help the company's CEO to make an important decision about future product lines. More precisely, students are asked to analyse data and decide which products the company should produce next year, in order to maximize its profit. To complete this activity, students need to use the Insight Toy Company data set ("Insight_Toy_Demo") provided by SAS Visual Analytics on Teradata University Network and then proceed with sharing and discussion of individual insights on the enterprise social media platform called Yammer. The overall learning activity was inspired by the well-known group-decision making process (Nunamaker, Denis, Valacich et al. 1991) that has been used by group decision support systems (GDSS) for many decades. In our case this process focused on cross-functional integration informed by insights obtained while using VA software.

*- Step 1: Data visualization using SAS VA (from TUN)*

For the purposes of this activity, students are divided into three groups (of two or more members), corresponding to three different functional areas/departments. Using the data set "Insight-Toy-Demo", their first task is to create three different visualisations (one per group) corresponding to three areas/departments: sales, finance and customer service.

More precisely, the sales department group needs to create the *Product Sale by Product Line* visualisation in SAS VA (Figure 2a). The finance group is given the task of creating the *Product Material Cost by Product Line* visualisation (Figure 2b). The third group is asked to create the *Customer Satisfaction by Product Line* visualisation (Figure 2c). Please note that after experimenting with different combinations, we came up with this particular set of visualisation, as all of them clearly show that the company is currently experiencing a business problem.

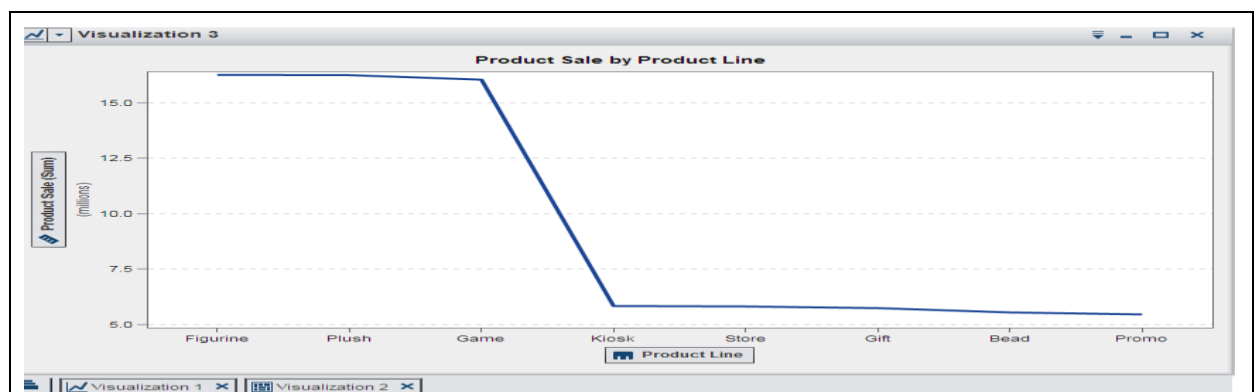

*Figure 2a.  Product Sales by Product Line Visualization*

In essence, students are provided with an incomplete functional data and asked to make a decision based on data they have. Even more, each visualisation was deliberately designed to lead students into selecting what is best for their allocated department but not necessarily for the other.

Please note that Figure 2a has two limitations, caused by the limitations of the provided data set. First, the reason for choosing the sales figure only is based on the (limited) assumption that sales figures serve as a preliminary indicator for profitability. Second, the current (teaching) version of SAS VA did not allow the user to add units or currency to the numbers which in this case will be $ for the Product Sale figures. The same applies to Figure 2b, where Product Material Cost should be interpreted to include $ currency.

Figure 2c shows customer satisfaction for each product line presented. This information alone might not provide a meaningful clue for product sales or profitability. However, the graph is designed based on the assumption that students assigned to the Customer Service department only have access to customer satisfaction data for their decision making in the first instance.

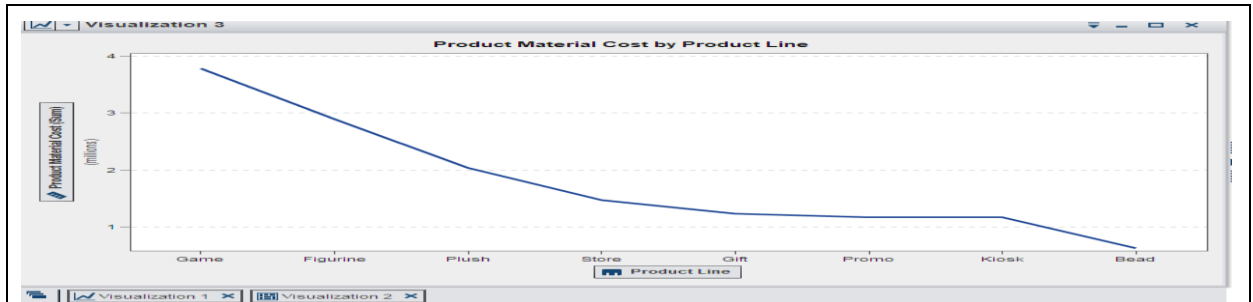

*Figure 2b. Product Material Cost by Product Line*

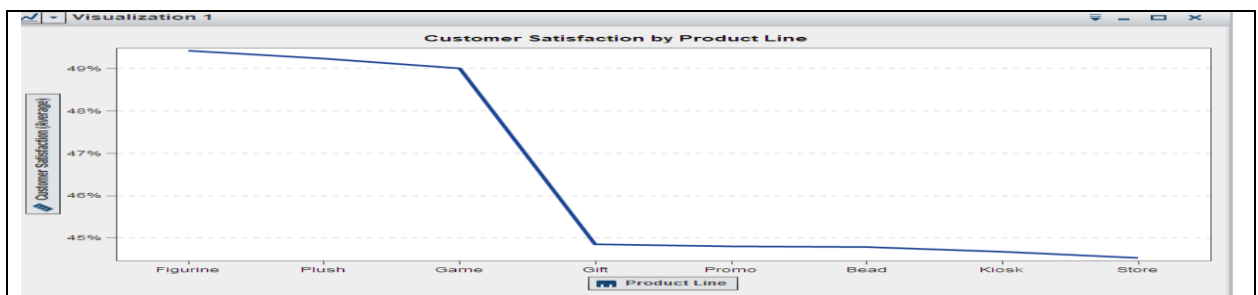

*Figure 2c. Customer Satisfaction by Product Line*

**- Step 2: Importing the resulting SAS-VA data visualisations into Yammer**

This is a "transition" task linking data visualisation and collaboration. The previous step resulted in three data visualisations that need to be saved as images (pdf) files in order to be imported into Yammer. To ensure the consistency of results as well as the same basis for Steps 3-6 for all students, this task is performed by the teacher who needs to upload the suggested (correct) solutions, as follows. First of all, the teacher will set up a private group on Yammer for each departmental group (finance, sales, customer service). Then s/he will upload the corresponding graph (image) into each group's collaborative space on Yammer (one per group).

**Step 3: Step 3: Collaborative activity based on a single (functional) perspective**

Each group is now required to analyse the graph assigned to their department, discuss their individual insights and select/comment on one or two products they all agree should be produced next year. Based on the existing "silo data" and limited "departmental perspectives" teams are very likely to select different items. For example, the finance team is likely to select the products that are less expensive to manufacture, while the customer service team is likely to be guided by customer satisfaction. While their discussions are supported by the collaborative tool (Yammer), students are encouraged to go back to SAS VA and further explore data related to their particular graph to gain additional insights.

**- Steps 4: Collaborative activity of exploring disciplinary perspectives, using internal and external data.**

While the previous step focused on a single (functional) perspective, this step challenges students to take into account the perspectives of other functional departments. After they discuss their insights and select their preferred product based on internal data (Step 4a), they will be given access to external data that would further challenge their decision-making process (Step 4b).

Thus Step 4 starts with the teacher noting the most popular products and product-related comments generated by each group in their respective collaborative space. The teacher will then use these observations to create a Yammer poll for all departments (all students) to vote. At this point, the groups are given all three pdf files generated in Step 1 of this activity, representing the cross-departmental data they need to take into account before casting their votes.

Even though the final consensus is achieved by voting (with all students having 'equal' votes), it is important to acknowledge that this is rarely the case in real-life organisations. Different organisational roles (e.g. senior-level executives) have the positional power as well as the responsibility for making the final decision. Therefore, they might be informed by the discussion shared on a social media platform, in order to get the multidisciplinary perspective before making their own decision. This particular point could be also incorporated into the learning activity, after student voting, to make them even more aware of the challenges included in collaborative decision making in a real context.

Now, suppose that most students vote for 'Plush' as it generates the highest sales and has the highest customer satisfaction, although it incurs a very high production cost. In Step 4b, the teacher will then introduce some external data on "Plush" and import it into Yammer. In this particular example, we used data from (i) Twitter (obtained by using #plush and #toy search); (ii) Google Trends (showing specific Google searches for "plush" toys per months and geographical locations); and (iii) news articles from local/regional/international digital newspapers. We used our leading national news web site because it published some highly relevant and timely articles about "unsafe plush toys" being sold by local department stores in Australia as well as the relevant articles published by the Wall Street Journal.

These external sources of data required students to re-consider their original selections (based on disciplinary perspectives). The same external data also allowed the teacher to ask additional questions about the chosen product. For example, based on the Google trend data (as shown by Figure 3), we could ask questions such as: "Which country should we export the Plush product to?" Which month would be the best for our promotional campaign?" This was also an opportunity to discuss limitations of these and other external sources of data, as they offer an additional but in many ways limited perspective (e.g. data from Google trends do not indicate future sales opportunities).

Similarly, using Twitter data it was possible to ask more questions. For example, "Based on a snapshot of Twitter data provided on Yammer, do you think that people still like, dislike or don't have any opinion (are neutral) about the plush toys? Note that Twitter data also offer an opportunity to use sentiment analysis to extend this activity even further. For example, the company may analyse its own Twitter data (currently not included into the provided data set). Interested BI&A educators may consider excellent examples of sentiment analysis by Mazzola, Kanat and Goul (2012).

The major limitation of Step 4 as presented here is that relevant social media data might not always be available for every product from the Insight Toy Company data set (from SAS-VA).

*- Step 5: Collaborating to translate VA-enabled insights into action*

After completing the previous steps over several learning sessions, in this step students are asked to go back to the original business question and write a memo to the company's CEO, providing a convincing data-supported argument for the chosen product. This could be done individually or in groups, with Yammer providing support for collaborative editing activities. Subject to available data and VA software capabilities, this step could be extended to include visual storytelling (i.e. telling a story with data) that could be presented to the teacher (acting as the CEO) or to an industry panel.

*- Step 6 – Reflection-on-action (individual and/or collaborative)*

The final step provides an opportunity for students to reflect on their learning experience, in particular the challenges of co-creating analytical insights across functional boundaries through collaboration, in particular the challenge of using internal and external data to create and refine analytical insights as they went through different steps. This reflective activity could be done individually or in groups, prompting further discussion, face-to-face or online.

This step is also an excellent opportunity to bring in the same patterns of challenges in using visual analytics found across industry sectors, as listed in the third section of this paper. As already stated, these patterns included data silos, different internal and external sources and so on. After going through steps 1-5 of this activity, students could relate to some, if not all of these challenges, based on their own experience.

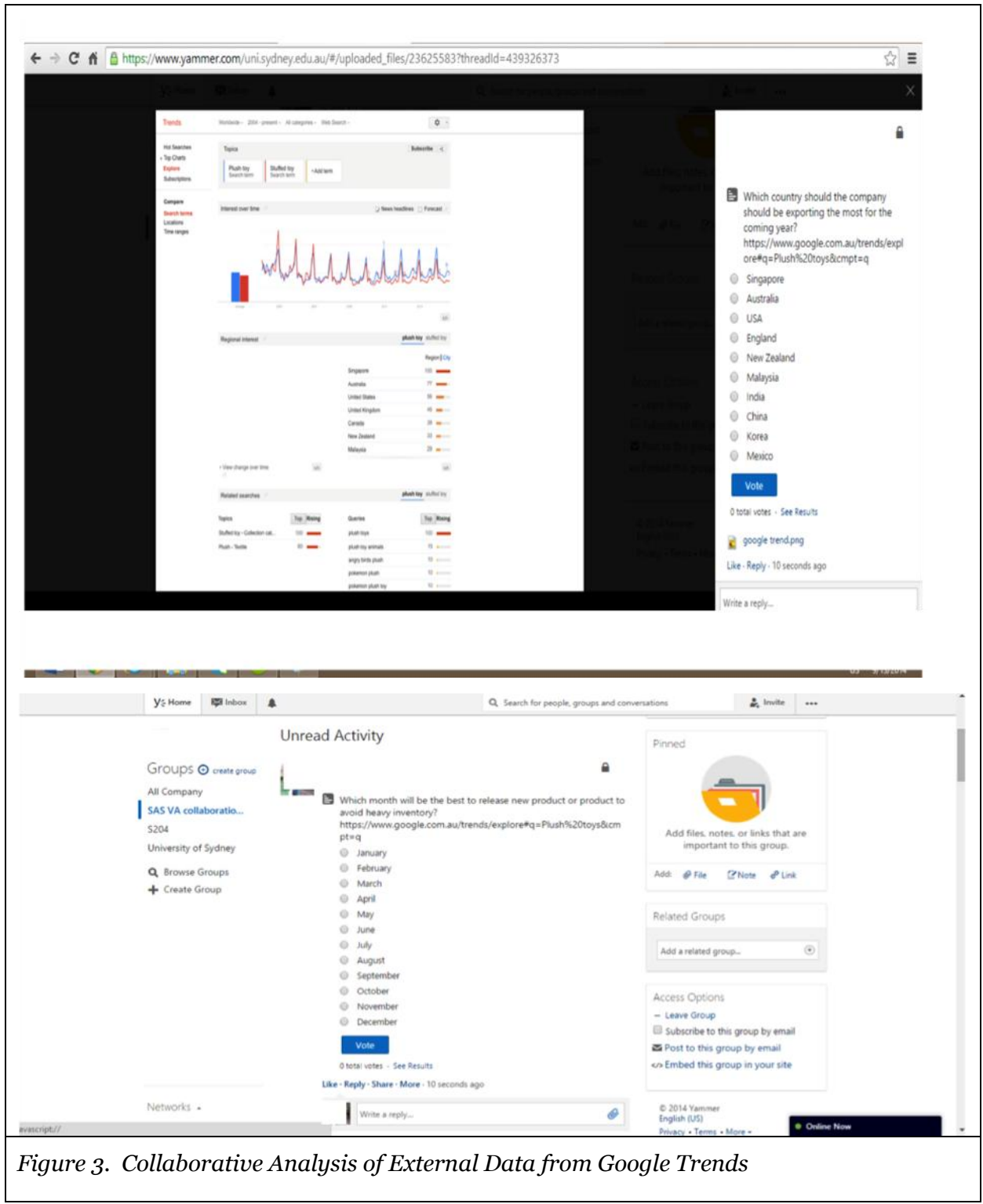

*Figure 3. Collaborative Analysis of External Data from Google Trends*

## 6  Discussion

The main objective of this paper was to illustrate an innovative learning activity that combines data visualisation and collaboration across functional data "silos". Apart from giving students hands-on experience with the two leading software platforms (SAS-VA and Yammer), this innovative activity makes a contribution to both industry and educational practices. For example, a very recent thought leadership white paper on collaboration by SAS (Bailey, et. al. 2014) offers further support for our idea of combining visualisation and collaboration. More precisely, it demonstrates how SAS VA could be combined with office productivity tools (such as MS Office) in order to enable better collaboration, deemed necessary to bring together often isolated (disciplinary) insights. We go one step further and propose to combine SAS VA with the leading enterprise social media platform made available to staff and students in our university environment.

From the theoretical perspective, the activity aims to illustrate the difference between *meaning-free* data (facts) held by VA (BI&A) systems and *information* which corresponds to different interpretations made by human decision-makers in their own context. As such, this information cannot be provided or stored by systems no matter how sophisticated they might be. This particular aim is very important, especially in the age of Big data and open data, with data and various data-processing algorithms increasingly seen as "objective" or even "authoritative". Consequently, they might be used to justify new management and societal practices of "datification", such as for example the one described in (Cecez-Kecmanovic and Marjanovic 2015). Students can also observe aadditional organisational complexities created when data sets generated in one context are propagated across different organisational boundaries and as such subjected to different interpretations when used in other contexts or even subjected to organisational politics.

Furthermore, when considered from the Big Data perspective, the activity illustrates large *volume* and *variety* of data. More precisely, the Insight Toy Company data set, as currently provided by SAS-VA, is made up of 3.5 million records and 60 columns, representing 14 years of Finance, Manufacturing and Sales & Marketing Data, collected from 127 facilities around the world (SAS-VA 2014). Furthermore, the social media data are by nature unstructured, and as such, could be taken to illustrate the variety component of Big data. However, the third component – velocity – is not included in this example as data are not accessed in a real-time mode. Instead, decision makers (students) were working with historical data, that although realistic, cannot be taken to represent real-time data. While real-time access to data is be possible (especially with social-media data), learning activities, such as the one described in this paper, need to be designed in advance with somewhat predictable data sets (in terms of their content), in order to scaffold student learning towards the intended learning objectives.

Furthermore, instead of Yammer that is an enterprise social media platform and therefore requires a site license, it would be possible to use other social media platforms such as Facebook. However, in our university environment any use of Facebook for teaching and learning purposes was not possible due to the current and unforeseen ethical consequences. For example, asking students to open new or use their private Facebook accounts for teaching and learning activities (especially group activities) is not considered to be ethical, due to lasting effects and possibly yet unforeseen future consequences, associated with this public social media platform. This may not be the case with other universities, but nevertheless should be taken into account by the educators.

Other possible options to support the collaborative part of this activity include more traditional computer-supported collaborative tools (including open source), groupware tools, learning management tools (such as the Whiteboard option on Blackboard), tagging software platforms or open web 2.0 tools (such as password-protected wikispaces). After exploring these options, we decided to go with the leading (corporate) enterprise social-media platform made available to all students and staff.

# 7   Conclusion, Limitations and Future Work

In this paper we respond to the challenge of creating realistic learning activities informed by leading industry practices, which is one of the key challenges of BI&A education today, as confirmed by Wixom et al. 2014). Rather than focusing on technical features of software platforms, in particular SAS-VA, and teaching a skilled-based course following user manuals, this activity shifts the focus to organisational challenges of VA in-use.

Ultimately, the process of turning data into insights and then into an action is not mechanistic as often portrayed in industry press. Instead, it is human-driven and human-centric. As demonstrated by this learning activity, and ultimately shown to students, this process is always complex and very much related to "human intelligence" (i.e. human individual and collective ability to make sense of data in a given context), rather than "business intelligence" provided by BI&A systems.

The activity presented in this paper has a number of limitations. At this point of time, the activity is designed around the *current* (teaching) version of SAS VA made available on TUN, including the existing data sets. In other words, we had to find a possible (realistic) business scenario around the existing set of data, without any additional option to upload more data sets into our version of SAS VA or change the data to fit our purpose. Consequently the individual steps had to be designed to "fit" the available data set, rather than the other way around.

However, going around this problem we allowed additional sets of data posted on Yammer, as demonstrated in Step 4. Obviously this is not ideal, as additional data files posted on corporate social media platforms further propagate the problem of data silos. At the same time this could lead to interesting classroom discussion about the consequences of doing something like it in a real-life environment.

Our future work includes design of new VA activities, based on the same or similar set of underlying patterns of VA-related challenges (as described in the paper). We are currently exploring other data visualisation platforms and possible data sets. Finally, our insights into practitioner stories (obtained through case studies, practitioner interviews and observations) have also inspired our current exploration of data-driven storytelling that we also plan to incorporate into future learning and teaching activities.

## Acknowledgments

This work was supported by the Australian Office for Teaching and Research (OLT) grant ID12-2407 *"Enhancing Collaborative Learning in Information Systems Business Analytics Using Data Visualisation and Manipulation Techniques"*.

The author would like to thank Ms. Thireindar Min, a research assistant employed by the OLT grant, for her help with the practical implementation of this learning activity.

## Copyright